\documentclass[11pt]{article}
\usepackage{moriond,epsfig}

\def\be{\begin{equation}}
\def\ee{\end{equation}}
\def\bea{\begin{eqnarray}}
\def\eea{\end{eqnarray}}

\begin{document}
\vspace*{4cm}
\title{THE DISCOVERY POTENTIAL OF SUPERSYMMETRY AT CMS WITHIN THE mSUGRA MODEL 
USING SAME-SIGN DI-MUONS}

\author{ \underline{D.I. FUTYAN} }

\address{Department of Physics, University of California, Riverside, CA 92521, USA}

\author{ A. DROZDETSKIY, D. ACOSTA, G. MITSELMAKHER }

\address{Department of Physics, University of Florida, Gainsville, FL 32611-8440 USA}

\maketitle\abstracts{
    A detailed study of the same-sign muon signature within the mSUGRA
    model is described. Selection criteria based on the missing
    transverse energy in the events and the jet and muon transverse
    momenta are applied. The results indicate that an excess of SUSY
    events over the Standard Model background processes can be
    statistically significant for an integrated luminosity of less than 
    $10 {\rm fb}^{-1}$ for many benchmark points with $m_{1/2}$ up to 
    $650$ GeV/$c^2$.
}
\section{Introduction}

The mSUGRA model is a popular simplification of the Minimal Supersymmetric 
Standard Model (MSSM)~\cite{PDGSUSY}, with only five free parameters:
$m_0, m_{1/2}, \tan\beta, A_0, {\rm sign}(\mu)$.  In this study the following 
values of the parameters were used:
${\rm sign}(\mu) > 0$, \mbox{$A_0 = 0$,} $\tan\beta = 10, 20, 35$ and 20
$(m_{1/2},m_0)$-points. All points were chosen so as to satisfy recent
theoretical and experimental constraints~\cite{ELLIS}.

The same-sign di-muon signature was chosen to evaluate the chances of
discovering SUSY, since it is very clean, has high trigger efficiency
and relatively small background. A recent published theoretical study of the
signature at the Tevatron can be found in Ref.~\cite{MATCHEV}.  

The aim of the present work was to investigate the region in the mSUGRA
parameter space which is accessible to CMS at the LHC start-up.

\section{Event Generation, Simulation and Reconstruction}

Coupling constants and cross sections in the Leading Order (LO)
approximation for SUSY processes were calculated with ISASUGRA
7.69~\cite{ISAJET}. Next to Leading Order (NLO) corrections were
calculated with PROSPINO~\cite{PROSPINO}.  Cross sections 
for SM processes were calculated
using PYTHIA 6.220~\cite{PYTHIA} and CompHEP 4.2p1~\cite{COMPHEP}. For
several SM processes (${\rm t\bar{t}, ZZ, Zb\bar{b}}$), the NLO
corrections are known and were used~\cite{VALERI}.  
Preselection cuts were applied at generator level for signal and background 
requiring at least two same-sign muons with $P_T > 10$ GeV and $|\eta| < 2.5$.

Full simulation of the CMS detector 
was performed using GEANT-based CMSIM~\cite{CMSSOFT}.  Data 
digitization and reconstruction were performed with the
ORCA~\cite{CMSSOFT} reconstruction package. Pile-up was not
taken into account in this study.  It was verified that all events 
satisfying the selection criteria passed both L1 and HLT muon triggers.

\section{Signal processes}
\label{sec:signal}

\begin{figure}
  \small
  \begin{center}
    \resizebox{11cm}{!}{\includegraphics{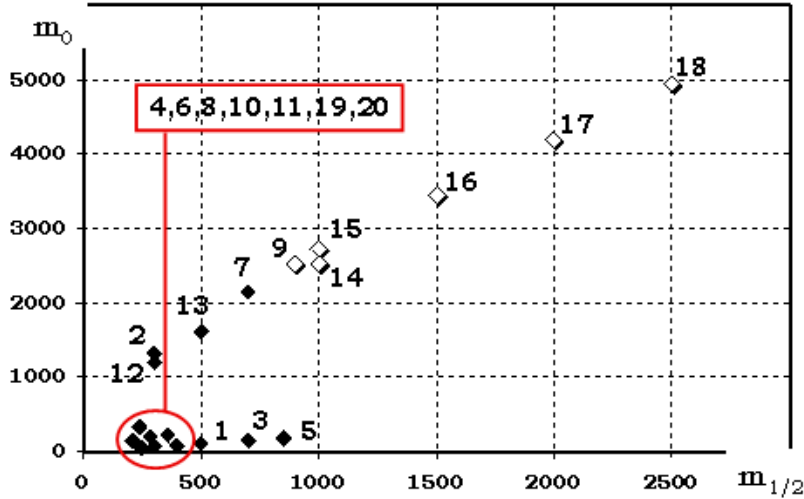}}
    \caption{The mSUGRA benchmark points investigated in this analysis.}
    \label{fig:points}
  \end{center}
\end{figure}

The 20 points chosen for this analysis are shown in
Fig.~\ref{fig:points}. Points 2, 11, 16, 19 are mSUGRA benchmark
points taken from Ref.~\cite{SUSYBENCH} (two of them, points 2 and 16,
were modified for a top-quark mass of 175 GeV as used for all other
points).

For each mSUGRA point, the LO cross section was calculated and hence
the number of events expected for an integrated luminosity of 10 ${\rm
fb}^{-1}$. NLO corrections were applied to the latter for all mSUGRA points.
The six points shown on the plot by empty markers (\# 9, 14-18) 
are found to have a cross section too small for a target integrated 
luminosity of $10 {\rm fb}^{-1}$ and are not considered further in the study.

\section{Background processes}

The cross sections and the number of generated and selected events for
several sources of background are listed in 
Table~\ref{tab:bckg1}. In the ${\rm tb}$, ${\rm tqb}$, ${\rm
\bar{t}b}$, ${\rm \bar{t}qb}$ processes, the top quark was forced to
decay to ${\rm Wb}$ and the ${\rm W}$ was forced to decay into a muon
and neutrino. In the
${\rm Zb\bar{b}}$ process, the ${\rm Z/\gamma^*}$ was forced to decay to
$\mu^+\mu^-$, and these muons were required to have an invariant mass larger 
than 5 GeV/$c^2$.

\begin{table}[htb]
  \small
  \caption{Cross sections and numbers of events for SM processes
  for an integrated luminosity of $10 {\rm fb}^{-1}$. $N_{\rm generated}$ 
  is the
  unweighted number of generated events, $N_{\rm selected}$ is the
  unweighted number of preselected events, $N_1$ is the number of
  events for an integrated luminosity of 10 ${\rm fb}^{-1}$, $N_2$ is
  the number of events after preselection cuts (at least two
  same-sign muons with $P_T > 10$GeV/c.)}
  \label{tab:bckg1}
  \begin{center}
    \begin{tabular}{|c|c|c|c|c|c|} \hline
      Process  & $\sigma, \: pb$ & $N_{\rm generated}$ & $N_{\rm selected}$ & N1 & N2 \\ \hline
      ${\rm tb}$ & 0.212 & 18,999 & 1,000 & 2,120 & 112  \\
      ${\rm tqb}$ & 5.17 & 28,730 & 1,000 & 51,700 & 1,798  \\
      ${\rm \bar{t}b}$ & 0.129 & 13,588 & 745 & 1,290 & 71  \\
      ${\rm \bar{t}qb}$ & 3.03 & 28,359 & 1,000 & 30,300 & 1,067  \\
      ${\rm ZZ}$ & 18(NLO) & 433,489 & 1,000 & 180,000 & 256  \\
      ${\rm ZW}$ & 26.2 & 368,477 & 1,000 & 262,000 & 727  \\
      ${\rm WW}$ & 26.2 & 894,923 & 41 & 702,000 & 39.7  \\
      ${\rm t\bar{t}}$ & 886(NLO) & 931,380 & 15,000 & 8,860,000 & 142,691  \\
      ${\rm Zb\bar{b}}$ & 232(NLO) & 359,352 & 2,000 & 2,320,000 & 12,924  \\
      $All$ &  &  &  &  & 160,000  \\ \hline
    \end{tabular}
  \end{center}
\end{table}

The contribution of several other potential background processes 
were investigated: WWW, ZWW, ZZW, ZZZ, WWWW, ZWWW, ZZWW, ZZZW, ZZZZ, 
${\rm t\bar{t}W, t\bar{t}Z, t\bar{t}WW, t\bar{t}ZW, t\bar{t}ZZ}$.
No detailed simulation was performed for these
processes. An estimation was obtained from the process cross section
(calculated with CompHEP) and the branching fraction into muons.
All but the ${\rm t\bar{t}W, t\bar{t}Z}$ processes were found to be 
negligible and are neglected in this analysis. The ${\rm t\bar{t}W}$ and 
${\rm t\bar{t}Z}$ backgrounds are also neglected here, but require 
further investigation.  QCD multi-jet production is a further possibly 
significant background which is yet to be investigated.

\section{Event Selection and cut optimization}

\begin{figure}[hbtp]
  \begin{center}
    \resizebox{9cm}{!}{\includegraphics{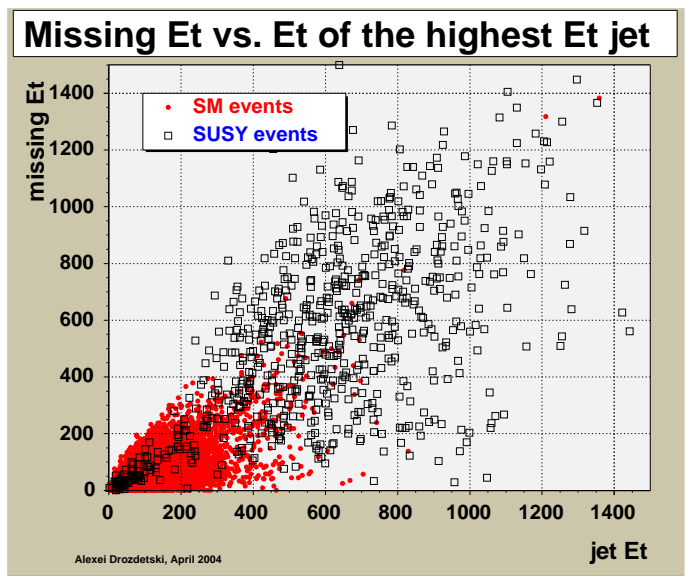}}
    \caption{Missing transverse energy vs. jet $E_T$ for SUSY (open
      squares) and SM events (points) after full simulation and
      reconstruction.}
    \label{fig:jetmet}
  \end{center}
\end{figure}

The distributions of kinematic variables such as missing $E_T$, jet
$E_T$ and muon $P_T$ are in general very different for SUSY and SM processes.  
Fig.~\ref{fig:jetmet} shows an example of these differences. Suitable cuts
on such variables can therefore be used to suppress the SM background.

Five discriminating variables were selected: the total missing transverse 
energy, the transverse energy of the highest and third highest $E_T$ jets, 
and the transverse momenta of the two muons. For each variable a
set of possible selection cuts was defined.  For each possible combination of 
these cuts for the five variables, the significance, the signal to 
background ratio and the number of expected events for $10 {\rm fb}^{-1}$ were 
calculated.  The significance was calculated using the following expression
\cite{BITYUKOV}: $S_{12} = 2 (\sqrt{N_S + N_B} - \sqrt{N_B})$, where
$N_S$ and $N_B$ are the total number of signal and background events, 
respectively, at the mSUGRA point in question. An iterative optimization 
procedure was used to identify sets of cuts for which these three quantities 
are close to maximum for all mSUGRA points.  The 
two sets of cuts shown in Table~\ref{tab:cutsfin} were together found to 
satisfy this requirement.

\begin{table}[htb]
  \small
  \caption{Chosen cut sets.  $E_{T, {\rm jet}_1}$ is the $E_T$ of
the highest $E_T$ jet, $E_{T, {\rm jet}_3}$ is the $E_T$ of the
third highest $E_T$ jet, $P_{T, \mu_1}$ and $P_{T, \mu_2}$ are the two
highest $P_T$ values of the same-sign muons.}
  \label{tab:cutsfin}
  \begin{center}
    \begin{tabular}{|c|c|c|c|c|c|} \hline
      set & miss. $E_T$, GeV & $E_{T, {\rm jet}_1}$, GeV & $E_{T, {\rm jet}_3}$, GeV & $P_{T, \mu_1}$, GeV & $P_{T, \mu_2}$, GeV  \\ \hline
      1 & $>$ 200  & $>$ 0 & $>$ 170 & $>$ 20 & $>$ 10 \\
      2 & $>$ 100  & $>$ 300 & $>$ 100 & $>$ 10 & $>$ 10  \\ \hline
    \end{tabular}
  \end{center}
\end{table}

\section{Results}

The expected number of signal and background events, the significance, and the 
signal to background ratio for each mSUGRA point studied are shown in 
Table~\ref{tab:results1}.

\begin{table}[htb]
  \small
  \caption{Expected number of events, significance and signal to background 
ratio after all cuts, for the 20 SUSY benchmark points considered. The ``SM'' 
row gives the expected number of the SM background events after all cuts for 
all considered processes.  The indices set1 and set2 refer to cut
sets \# 1 and 2 respectively.}
  \label{tab:results1}
  \begin{center}
    \begin{tabular}{|c|c|c|c|c|c|c|} \hline
         & $N_{{\rm set}1}$ & $S_{12,{\rm set}1}$ & $S/B_{{\rm set}1}$ & $N_{{\rm set}2}$ & $S_{12,{\rm set}2}$ & $S/B_{{\rm set}2}$  \\ \hline 
      SM & 69.5$\pm$6.0 &  &  & 432$\pm$8.8 &  &   \\
      1  & 95.9$\pm$6.7 & 9.05 & 1.38 & 184$\pm$9.3 & 8.06 & 0.43 \\
      2  & 282$\pm$20 & 20.8 & 4.06 & 560$\pm$29 & 21.4 & 1.3 \\
      3  & 17.7$\pm$1.1 & 2 & 0.25 & 30.4$\pm$1.4 & 1.44 & 0.07 \\
      4  & 365$\pm$73 & 25 & 5.26 & 1590$\pm$152 & 48.4 & 3.7 \\
      5  & 6.54$\pm$0.37 & 0.77 & 0.094 & 9.6$\pm$0.45 & 0.46 & 0.002 \\
      6  & 277$\pm$35 & 20.6 & 4.0 & 1030$\pm$67 & 35 & 2.4 \\
      7  & 6.7$\pm$0.35 & 0.78 & 0.096 & 8.31$\pm$0.39 & 0.4 & 0.019 \\
      8  & 188$\pm$17 & 15.5 & 2.71 & 530$\pm$28 & 20.5 & 1.2 \\
      10 & 515$\pm$78 & 31.7 & 7.41 & 1950$\pm$151 & 56.1 & 4.5 \\
      11 & 137$\pm$11 & 12.1 & 1.98 & 322$\pm$18 & 13.4 & 0.75 \\
      12 & 409$\pm$30 & 27.1 & 5.89 & 781$\pm$42 & 28.1 & 1.8 \\
      13 & 58.8$\pm$3.3 & 6 & 0.85 & 86.9$\pm$4 & 4 & 0.2 \\
      19 & 377$\pm$59 & 26.5 & 5.43 & 1220$\pm$106 & 39.8 & 2.8 \\
      20 & 279$\pm$36 & 20.6 & 4.01 & 996$\pm$67 & 34 & 2.3 \\ \hline
    \end{tabular}
  \end{center}
\end{table}

In addition to the six points already excluded due to low cross section 
(see Section~\ref{sec:signal}), three further points are out of reach 
(significance less than five) for $10{\rm fb}^{-1}$ of integrated luminosity 
for both cut sets: points 3, 5 and 7. For the benchmark mSUGRA points with
significance greater than five the signal to background ratios are
greater than 0.4.

Referring to Figure~\ref{fig:points}, with points 3, 5 and 7 additionally 
excluded, an approximate sensitive area for $10 {\rm fb}^{-1}$ can be 
defined on the $m_{1/2}$ parameter axis as $m_{1/2} < 650$. In order to put 
bounds also on $m_0$ it is necessary to investigate more
mSUGRA benchmark points with $m_{1/2} < 650$ GeV and $m_0 > 1500$ GeV.

\section{First estimate of systematic effects}

The stability of the significance with respect to the uncertainty on
the signal acceptance and the background normalization was verified. A
correlated variation of the SM event number ($+30\%$) and expected
number of SUSY events ($-30\%$) was applied.  It was found that
the significance of only one mSUGRA point (\# 13) 
drops below the discovery level.

\section{Summary}

A detailed study of the same-sign muon signature arising from 20
different points in the mSUGRA model was performed. The region in the
$(m_{1/2},m_0)$-plane with $m_{1/2} < 650$ GeV$/c^2$ is accessible by
the CMS at $5\sigma$ level at the LHC start-up ($10 {\rm fb}^{-1}$).

\end{document}